\documentclass[%
 reprint,
 amsmath,amssymb,
 aps,
]{revtex4-1}

\usepackage{subfigure}
\usepackage{graphicx}
\usepackage{dcolumn}
\usepackage{bm}
\usepackage{fourier}

\begin{document}


\title{The propagation and transformation of light beam on curved surface}

\author{ZIQIU SHAO}
 \altaffiliation{Zhejiang Province Key Laboratory of Quantum Technology and Device, Department of Physics, Zhejiang University, Hangzhou 310027, China}
\author{ZHAOYING WANG}%
 \email{zhaoyingwang@zju.edu.cn}
\affiliation{Zhejiang Province Key Laboratory of Quantum Technology and Device, Department of Physics, Zhejiang University, Hangzhou 310027, China}%

\date{\today}

\begin{abstract}
Starting from the wave function with abstract index notations, with the help of Wentzel–Kramers–Brillouin (WKB) approximation, a generalized coordinate-independent expression for the evolution of light beam on curved space is derived. By defining the propagation axes, the expression reduces to integrable Green functions. As a first application, one Gaussian beam on curved surface is analyzed for its propagation and transformation. The stationary status and Talbot effect of light field on different shapes of curved surfaces are discussed. We also notice that a periodical fractional Fourier transform can be perfectly achieved during the free propagation on curved surface, which must be implemented by additional optical elements in flat case. We hope our theoretical results can provide some references for the practical application in curved space.
\end{abstract}

\pacs{Valid PACS appear here}
\maketitle


\section{Introduction}
The characteristics of light beam propagation upon curved surface is now drawing increasing attention. It contains a vast range of applications towards versatile topics. In astrophysics, the Hanbury Brown and Twiss (HBT) measurement is a method to examine the angular size of stars, and it can be used to evaluate the curvature of surface\cite{RN15}\cite{RN39}. Similarly, Wolf effect is developed on curved surface, which could also be a method of surface shape evaluating\cite{RN24}. Surface Plasmon Polaritons (SPPs) are naturally surface bounded wave, for non-planer surface, the dynamics of it obey the generalized wave function along with space metric\cite{RN25}. The equivalence between the metric of surface and the index gradient of waveguide was discussed through conformal mapping as well, one subtle example is the comparison between Maxwell’s fisheye and the sphere\cite{RN42,RN26,RN10,RN14}. For classical optics, as the wave length of optical beams is tiny when compared with macroscopic scale, it is acceptable to treat the light transmission on curved surface under the frame of geometrical optics\cite{RN26,RN10,RN11,RN2,RN28}. But with the increasing research demand of microstructure whose spatial size is of the same order of magnitude with respect to the wavelength, the field on it must be treated by the means of wave optics. It is nowadays a widely discussed topic, wave optics on the curved space were theoretically revealed\cite{RN7,RN16,RN20,RN27,RN4,RN13,RN19,RN21,RN5} and experimentally observed\cite{RN8,RN23,RN43}.

But until now, there has not been a proper method of making connection between geometric optics and wave optics on curved surface. One may purse a differential geometric method for geometrical optics then turn to a Sturm-Liouville (S-L) eigenvalue problem for wave optics, separately. In classical flat space cases, a widely used method of connecting these two concepts is the WKB approximation\cite{RN44}, which is a mathematical processing of neglecting over high order derivative term(s) of amplitude in wave equation. The WKB approximation is a treatment upon flat manifold originally, to expand its usage to generalized manifolds, some efforts should be made. Two inspiring examples are the assumptions of surface of revolution\cite{RN4} and the radially symmetric media\cite{RN13}. The basic method is to take the effect of space curvature into the derivative operator of wave function, in consideration of proper-length transformation and eigenvalue condition, respectively.  However, these considerations would not be available near the poles of rotational symmetric system, limited with the discontinuity of spherical or polar coordinate at poles. What’s more, the currently used method allows non-geodesic lines to be the propagation axes, which is not tally with the facts and it is inevitable if adopting WKB approximation after defining specific coordinate variables in wave equation.

A new thought for this topic is to evaluate the effect of specific surface curvature after having solved the wave equation. A generalized WKB approximation is still utilized but before the definition of specific coordinate system. By defining the propagation axes, we derived integrable Green functions for light beams. It’s a concise and effective interpretation for the propagating field on curved surface. Our theory is confirmed by the flat space limit. The revolution of typical Gaussian beam is demonstrated thoroughly, a stationary state and the Talbot effect of light field are revealed. For a brand-new discovery, we found that a fractional Fourier transform can be perfectly performed on the surface of revolution without any additional optical elements, which may provide potential applications in signal processing. Finally, as an outlook, due to the complete description of optical phase in our method, the evolution of ultrashort pulses on curved surface can be readily further evaluated.

\section{Wave function on curved surface}
\label{sec:2}

A surface can always be treated as a hypersurface imbedded into a background manifold. 
Starting from pseudo-Riemannian space, the wave equation written with abstract indices gives\cite{RN45},
\begin{equation}\label{eq1}
{{\partial }^{a}}{{\partial }_{a}}{{A}_{b}}=-4\pi {{J}_{b}}
\end{equation}
where super- and sub-script obey the Einstein summation convention, that when an index variable appears twice in a single term and is not otherwise defined, it implies summation of that term over all the values of the index. 
The index $ a $ goes through over four independent coordinates $\left( {{q}_{1}},{{q}_{2}},{{q}_{3}},{{c}^{-1}}t \right)$.
${{J}_{b}}=\left( c\rho ,\mathbf{J} \right)$ is the covariant electric current vector, where $ \rho $ is the electric charge density and $ \mathbf{J} $ is the electric current vector.
$ A_b $ is the covariant vector potential. 
For two-dimensional surface, taking the separation\cite{RN1} with respect to the tangential part and the normal part $ A\left( {{q}_{1}},{{q}_{2}},{{q}_{3}},t \right)={{A}^{T}}\left( {{q}_{1}},{{q}_{2}},t \right){{A}^{N}}\left( {{q}_{3}},t \right) $, accompanying with the thin layer assumption, the wave function on two-dimensional sub-manifold writes:
\begin{equation}\label{eq2}
{{\partial }^{a}}{{\partial }_{a}}{{A}^{T}}_{b}+\left( {{H}^{2}}-K \right){{A}^{T}}_{b}=-4\pi {{J}^{T}}_{b}
\end{equation}
where $ H $ is the mean curvature while $ K $ is the Gaussian curvature.
After the separation, the index $ a $ now runs over $\left( {{q}_{1}},{{q}_{2}},{{c}^{-1}}t \right)$.
For macroscopic radii of curvature, the influence of the left second term of Eq.(\ref{eq2}) can be neglected.
By using the relationship between the field and vector potential and neglecting the polarization effects, the wave equation for propagating field on two-dimensional surface can be simplified as follows,
\begin{equation}\label{eq3}
{{\partial }^{a}}{{\partial }_{a}}{{E}^{T}}\left( t \right)-{{c}^{-2}}\partial _{t}^{2}{{E}^{T}}\left( t \right)=0
\end{equation}
with the index $ a $ now runs over $\left( {{q}_{1}},{{q}_{2}} \right)$.
Use the ansatz $ E=C{{e}^{ikL}} $ , substitute it back to Eq.(\ref{eq3})
\begin{equation}\label{eq4}
\begin{split}
&{{\partial }^{a}}{{\partial }_{a}}C+ik\left( {{\partial }^{a}}C{{\partial }_{a}}L+{{\partial }^{a}}L{{\partial }_{a}}C+C{{\partial }^{a}}{{\partial }_{a}}L \right)\\
&+{{k}^{2}}\left( 1-{{\partial }^{a}}L{{\partial }_{a}}L \right)C=0
\end{split}
\end{equation} 

Under the WKB approximation which works on trivial manifolds, the second derivation term should be neglected. Extract the real part and imagine part of Eq.(\ref{eq4}), we obtain the following expressions.
\begin{equation}\label{eq5}
\begin{matrix}
1-{{\partial }^{a}}L{{\partial }_{a}}L=0 \\ 
{{\partial }^{a}}C{{\partial }_{a}}L+{{\partial }^{a}}L{{\partial }_{a}}C+C{{\partial }^{a}}{{\partial }_{a}}L=0 
\end{matrix}		
\end{equation}

Here we should emphasize, until now, in our analysis, we haven’t specially referred to the metric of surface, and coordinates $\left( {{q}_{1}},{{q}_{2}} \right)$ remains undefined.
In Eq.(\ref{eq5}), it is always possible to define the first coordinate ${{q}_{1}}=L$ itself and the other coordinate perpendicular to it, writes ${{\partial }_{{{q}_{2}}}}L=0$.
The amplitude term $ C $ then satisfies
\begin{equation}\label{eq6}
4{{\partial }_{{{q}_{1}}}}C+C{{g}^{-1}}{{\partial }_{{{q}_{1}}}}g=0
\end{equation}
where $ g $ is the determinant of space metric upon the argument $ q_1 $ and $ q_2 $.
$ q_1 $ or $ L $ in Eq.(\ref{eq5}) was named the eikonal in flat space, we follow this note on curved surface.
It is a key in this section to connect the optical phase with the definition of curved coordinate system. For a two-dimensional surface, this eikonal can be defined as the length of trajectory curve. 
Utilizing the principle of Green function, the distribution of propagation field will be derived as follows. 
All the ingredients in Eq.(\ref{eq7}) now have a unambiguous physical correspondence.
\begin{equation}\label{eq7}
{{E}_{output}}={{E}_{incident}}\otimes C{{e}^{ikL}}
\end{equation}

\section{The evaluation of eikonal $ L $}

\begin{figure}[htbp]
	\centering
	\includegraphics[width=0.95\linewidth]{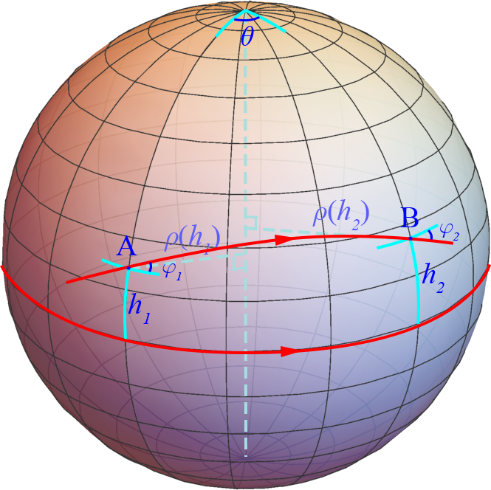}
	\caption{Coordinate system definition of surface of revolution. Red arrowed lines indicate the vector of wave field.}
	\label{fig1}
\end{figure}

A common and useful type of surface is called surface of revolution whose space metric writes $ d{{s}^{2}}=d{{h}^{2}}+\rho {{\left( h \right)}^{2}}d{{\theta }^{2}} $.
Drawn as sphere but without loss of generality, the sketch of surface and the corresponding coordinate system grids are plotted in Fig.\ref{fig1}.
The propagation trajectories are denoted as the red arrows, with $ A $ to be the incident point and $ B $ the exit point.
$ \theta $ is the angle of rotation, $ h $ is the arc length from the alternative point towards the maximum rotational circuit, and $\rho \left( h \right)$ is the radius of surface toward the rotation axis.
For a specific surface with constant Gaussian curvature, the reliance between $ \rho $ and $ h $ gives
\begin{equation}\label{eq8}
\rho \left( h \right)={{r}_{0}}\cos \left( h/r \right)
\end{equation}

For $r={{r}_{0}}$ one gets the sphere, for $r>{{r}_{0}}$ a spindle type, and for $r<{{r}_{0}}$ a bulge type surface. 
Once the space metric is given, the geodesic ray function or the optical trajectory writes
\begin{equation}\label{eq9}
\begin{split}
d\theta &=\frac{constdh}{\rho \left( h \right)\sqrt{{{\rho }^{2}}\left( h \right)-cons{{t}^{2}}}}\\
&=\frac{{{r}_{0}}\cos \psi dh}{\rho \left( h \right)\sqrt{{{\rho }^{2}}\left( h \right)-{{r}_{0}}^{2}{{\cos }^{2}}\psi }}
\end{split}
\end{equation}
where the constant parameter $const={{r}_{0}}\cos \left( h/r \right)\cos \varphi ={{r}_{0}}\cos \psi $ determines the trajectory of light ray.
In this parametrization, we are now able to complete the integration $ \theta $.
\begin{equation}\label{eq10}
\theta =-\frac{r}{{{r}_{0}}}\arcsin \frac{\tan {{h}_{2}}/r}{\tan \psi }+\frac{r}{{{r}_{0}}}\arcsin \frac{\tan {{h}_{1}}/r}{\tan \psi }
\end{equation}

The eikonal function $ L $ is the arc length between point $ A $ and $ B $, which could be calculated with the help of metric.
\begin{equation}\label{eq11}
L=\int_{{{h}_{1}}}^{{{h}_{2}}}{ds}=-r\arcsin \frac{\sin {{h}_{2}}/r}{\sin \psi }+r\arcsin \frac{\sin {{h}_{1}}/r}{\sin \psi }
\end{equation}

For the paraxial optical system with the propagation axis to be the $ h $ direction and starts at $ h=0 $, the constant now satisfies ${{r}_{0}}\cos \psi ={{r}_{0}}\cos \left( {{h}_{1}}/r \right)\cos {{\varphi }_{1}}={{r}_{0}}\cos {{\varphi }_{1}}$.
For the paraxial optical system with the propagation axis to be the $ \theta $ direction, eikonal $ L $ could be expanded with the axial length ${{r}_{0}}\theta $ and a deviation term, which is expressed as follows.
\begin{equation*}
\begin{split}
L&={{r}_{0}}\theta -\frac{r\tan \frac{{{h}_{2}}}{r}}{\sin {{\varphi }_{2}}}\left( \frac{1}{\cos {{\varphi }_{2}}}-1 \right)+\frac{r\tan \frac{{{h}_{1}}}{r}}{\sin {{\varphi }_{1}}}\left( \frac{1}{\cos {{\varphi }_{1}}}-1 \right)\\
&+O{{\left( \frac{{{h}_{2}}}{r} \right)}^{3}} +O {{\left( \frac{{{h}_{1}}}{r} \right)}^{3}}  \\ 
&={{r}_{0}}\theta -\frac{{{h}_{2}}{{\varphi }_{2}}}{2}+\frac{{{h}_{1}}{{\varphi }_{1}}}{2}+O {{\left( \frac{{{h}_{2}}}{r} \right)}^{3}} +O {{\left( \frac{{{h}_{1}}}{r} \right)}^{3}}   
\end{split}
\end{equation*}

With the relationship between incident(output) angles and the trajectory parameters,
\begin{align*}
& {{\varphi }_{1}}=\tan \frac{{{h}_{2}}}{r}\csc \frac{{{r}_{0}}}{r}\theta -\tan \frac{{{h}_{1}}}{r}\cot \frac{{{r}_{0}}}{r}\theta  \\ 
& {{\varphi }_{2}}=\tan \frac{{{h}_{1}}}{r}\csc \frac{{{r}_{0}}}{r}\theta -\tan \frac{{{h}_{2}}}{r}\cot \frac{{{r}_{0}}}{r}\theta   
\end{align*}
the eikonal function finally gets
\begin{equation}\label{eq13}
L={{r}_{0}}\theta +\frac{1}{2r}\left( -2{{h}_{1}}{{h}_{2}}\csc \frac{{{r}_{0}}\theta }{r}+{{h}_{1}}^{2}\cot \frac{{{r}_{0}}\theta }{r}+{{h}_{2}}^{2}\cot \frac{{{r}_{0}}\theta }{r} \right)
\end{equation}

The Eikonal function Eq.(\ref{eq13}) is the main result of this paper. Regardless of the reliance of eikonal definition on coordinate system, this calculation is flexible in conformal mapping on curved space. We are now able to describe several intuitive properties of the light propagation on two-dimensional curved surface.

For flat surface limit when $ r $ and $ r_0 $ approaches infinity, it’s able to define ${{r}_{0}}\theta =z$ as the propagation distance. 
${{r}_{0}}\theta /r$ remains a small angle, then the eikonal function Eq.(\ref{eq13}) can be simplified as
\begin{equation}\label{eq14}
L=z+\frac{{{\left( {{h}_{1}}-{{h}_{2}} \right)}^{2}}}{2z}
\end{equation}
which is the eikonal function under paraxial approximation in flat surface.

For a more complete description, the variation of amplitude term is also investigated. 
From Eq.(\ref{eq6}), the amplitude term can also be defined simply from the determinate of space metric, or equivalently speaking, the Gaussian (intrinsic) curvature.
After a little algebra, for surface with constant Gaussian curvature $K=1/{{r}^{2}}$ defined in this paper, the amplitude term gives $C=1/\sqrt{r\sin \left( L/r \right)}$.
When choosing the equator to be the propagation axis, it can be naturally rewritten as $ C=1/\sqrt{r\sin \left( {{r}_{0}}\theta /r \right)} $.

\section{Field propagation}

\begin{figure}[htbp]
	\centering
	\subfigure[]{
		\begin{minipage}[t]{0.45\linewidth}
			\includegraphics[width=\textwidth]{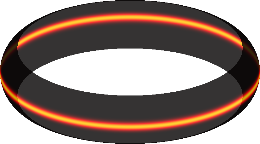}
		\end{minipage}%
		\label{2a}
	}
	\subfigure[]{
		\begin{minipage}[t]{0.5\linewidth}
			\includegraphics[width=\textwidth]{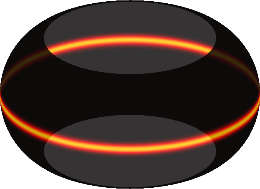}
		\end{minipage}%
		\label{2b}
	}
	\subfigure[]{
		\begin{minipage}[t]{0.33\linewidth}
			\includegraphics[width=\textwidth]{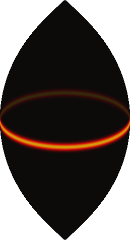}
		\end{minipage}%
		\label{2c}
	}
	\subfigure[]{
		\begin{minipage}[t]{0.28\linewidth}
			\includegraphics[width=\textwidth]{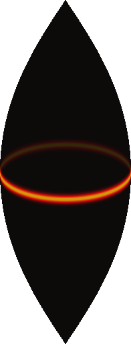}
		\end{minipage}%
		\label{2d}
	}
	\caption{The stationary Gaussian light beam with the size ${{\sigma }_{0}}=\sqrt{2r/k}$ on surface of revolution. 
		The parameters of surfaces are (a)$r=0.5{{r}_{0}}$, (b)$r=0.75{{r}_{0}}$, (c)$r=1.5{{r}_{0}}$ and (d)$r=2{{r}_{0}}$, respectively. }
	\label{fig2}
\end{figure}

\begin{figure}[htbp]
	\centering
	\subfigure[]{
		\begin{minipage}[t]{0.45\linewidth}
			\includegraphics[width=\textwidth]{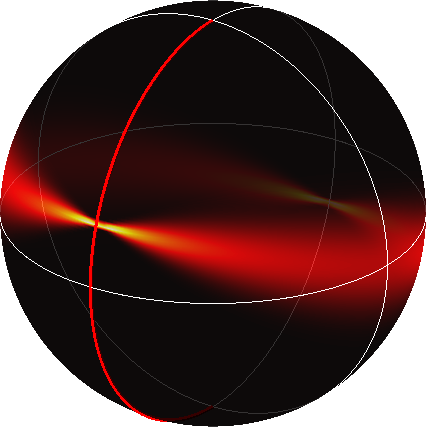}
		\end{minipage}%
		\label{3a}
	}
	\subfigure[]{
		\begin{minipage}[t]{0.45\linewidth}
			\includegraphics[width=\textwidth]{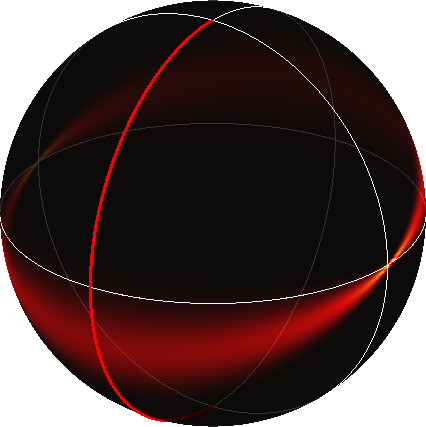}
		\end{minipage}%
		\label{3b}
	}
	\caption{The propagations of light beam on sphere. The initial incident positions are noted in red, with incident field as a Gaussian shape whose initial beam size (a)${{\sigma }_{0}}=0.2\sqrt{2r/k}$ and (b)${{\sigma }_{0}}=5\sqrt{2r/k}$. The Talbot effect property holds for alternative incident field. }
	\label{fig3}
\end{figure}

\begin{figure}[htbp]
	\centering
	\subfigure[]{
		\begin{minipage}[t]{0.6\linewidth}
			\includegraphics[width=\textwidth]{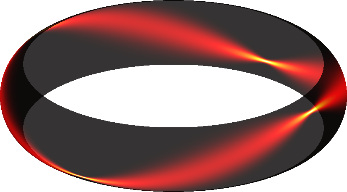}
		\end{minipage}%
		\label{4a}
	}
	\subfigure[]{
		\begin{minipage}[t]{0.3\linewidth}
			\includegraphics[width=\textwidth]{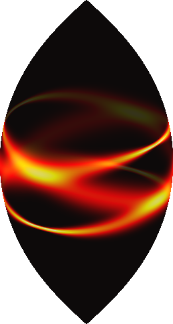}
		\end{minipage}%
		\label{4b}
	}
	\caption{Two typical examples of stationary solutions on the surface of revolution, (a)$r=0.5{{r}_{0}}$ and (b)$r=1.5{{r}_{0}}$. It shows that the translational symmetric circuit number are (a)$m=0.5$ and (b)$m=1.5$, respectively, indicating that the field will retrieve back to the incident form after $m$ circles of traveling. }
	\label{fig4}
\end{figure}

In laboratory, a most common and practical propagation field is the Gaussian beam, writes $E\left( {{h}_{1}} \right)={{E}_{0}}\exp \left( -{{h}_{1}}^{2}/{{\sigma }_{0}}^{2} \right)$, where $ \sigma_{0} $ is the initial beam size.
On the basis of Eq.(\ref{eq7}) and Eq.(\ref{eq13}), an analytic expression of propagation field with Gaussian shape on curved surface is deduced
\begin{equation}\label{eq15}
\begin{split}
& E\left( {{h}_{2}} \right)={{E}_{0}}\sqrt{\frac{{{\sigma }_{0}}}{{{\sigma }_{\theta }}}}\exp \left( -\frac{{{h}_{2}}^{2}}{{{\sigma }_{\theta }}^{2}} \right){{\operatorname{e}}^{i{{\phi }_{1}}}}{{\operatorname{e}}^{i{{\phi }_{2}}}}{{\operatorname{e}}^{ik{{r}_{0}}\theta }} \\ 
& {{\sigma }_{\theta }}=\sqrt{\frac{4{{r}^{2}}}{{{k}^{2}}\sigma _{0}^{\text{2}}}{{\sin }^{2}}\frac{{{r}_{0}}}{r}\theta +\sigma _{0}^{\text{2}}{{\cos }^{2}}\frac{{{r}_{0}}}{r}\theta } \\ 
& {{\phi }_{1}}=\frac{1}{2}\arctan \left( \frac{k\sigma _{0}^{2}\tan {{r}_{0}}\theta /r}{2r} \right) \\ 
& {{\phi }_{2}}=-\frac{\left( 4{{r}^{2}}-{{k}^{2}}{{\sigma }_{0}}^{4} \right)\tan {{r}_{0}}\theta /r}{4{{r}^{2}}{{\tan }^{2}}{{r}_{0}}\theta /r+{{k}^{2}}{{\sigma }_{0}}^{4}}{{h}_{2}}^{2}  
\end{split}		
\end{equation}	
$\sigma_\theta$ is the beam size at a propagation angle $\theta$.
${{\phi }_{1}}$ is the Gouy phase resulted from the transverse compression\cite{RN46}.
${{\phi }_{2}}$ is also a propagation phase, which describes a bending of wavefront.
The revolution of not only the illumination but also the optical phase on curved surface is revealed.

A special case of the initial beam size is ${{\sigma }_{0}}=\sqrt{2r/k}$. For this type of incidence, no matter how the surface shape with constant-Gaussian-curvature is, the beam size will keep a constant ${{\sigma }_{\theta }}={{\sigma }_{0}}$ during the propagation and the amplitude of beams remains unaltered, as shown in Fig.\ref{fig2}.
The stationary solution only depends on the relationship between the wave number and the Gaussian(intrinsic) curvature and have nothing to do with the extrinsic curvature, namely how this surface is embedded in background manifold.
The figures are plotted with a certain amount of transparency to show the revolution properties of field.
The wave number in these figures are set to be $k=1000/{{r}_{0}}$ when the variation of beam size is observable with respect to the radius of curvature.
On the other aspect of understanding, $k{{\sigma }_{0}}^{2}/2$ is the Rayleigh(diffraction) length of Gaussian beam, while $r$ is the principle radius of surface, the propagation stabilization requires the equivalence between them.
At the situation of stable transmission, the phase on axis varies as ${{\phi }_{axial}}=\left( 1/2r+k \right){{r}_{0}}\theta $, which principally explores a Gouy phase shift beyond an ordinary propagation phase.
Different from a logarithm reliance on flat space, the Gouy phase now behaves a linear superposition over propagation distance.

Another special case is the beam propagation on sphere (Fig.\ref{fig3}).
The sphere is a kind of surface with perfect isotropy, and this characteristic is inherited by the light field on it.
No matter how the incident field is, it will return to the initial status after one circle of traveling.
This corresponds to the fact that the spherical harmonic function is a complete set on sphere, any type of field could be expanded as a linear superposition of it.
The Talbot effect is clearly revealed in Fig.\ref{fig3} that the field diffuses and focuses spontaneously, and this property holds for alternative incident field.
From this Talbot effect, we find that the eigenvalue condition is contained naturally in the geometric of curved surface.
As proved in geometrical optics\cite{RN28} and now in wave optics as a major progress, the surface of revolution which satisfies $r=m{{r}_{0}}$ will always show its translational symmetry after going $m$ circles.
Two typical stationary solutions with limited circles are plotted in Fig.\ref{4a} for $r=0.5{{r}_{0}}$ and Fig.\ref{4b} for $r=1.5{{r}_{0}}$.
The self-interference effect is neglected in our simulation.

A more subtle feature of the curved surface is its possibility of performing a fractional Fourier transformation.
The emergent field $E\left( {{h}_{2}} \right)$ with an incidence $E\left( {{h}_{1}} \right)$ after a propagation angle $\theta$ gives
\begin{equation}\label{eq16}
\begin{split}
& E\left( {{h}_{2}} \right)=\frac{\exp \left( ik{{r}_{0}}\theta  \right)}{\sqrt{r\sin \left( {{r}_{0}}\theta /r \right)}}\exp \left( \frac{ik{{h}_{2}}^{2}}{2r}\cot \frac{{{r}_{0}}\theta }{r} \right)\times  \\ 
& \int{\exp \left( \frac{ik{{h}_{1}}^{2}}{2r}\cot \frac{{{r}_{0}}\theta }{r}-\frac{ik{{h}_{1}}{{h}_{2}}}{r\sin \left( {{r}_{0}}\theta /r \right)} \right)E\left( {{h}_{1}} \right)d{{h}_{1}}}  
\end{split}		
\end{equation}

\begin{figure}[htbp]
	\centering
	\includegraphics[width=0.95\linewidth]{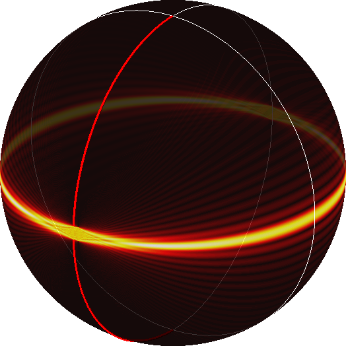}
	\caption{The diffraction pattern of curved surface with a plane incidence.}
	\label{fig5}
\end{figure}

On the other hand, the definition of fractional Fourier transformation\cite{RN29}
\begin{equation}\label{eq17}
\begin{split}
& {{\text{F}}_{\alpha }}f\left( x \right)=\frac{\exp \left( \frac{i\pi -2i\alpha }{4} \right)}{\sqrt{2\pi \sin \alpha }}\exp \left( -\frac{i{{x}^{2}}}{2}\cot \alpha  \right)\times \\
& \int{\exp \left( -\frac{i{{x}^{\prime }}^{2}}{2}\cot \alpha +\frac{ix{{x}^{\prime }}}{\sin \alpha } \right)f\left( {{x}^{\prime }} \right)d{{x}^{\prime }}}
\end{split}
\end{equation}
By means of the substitution ${{x}^{\prime }}={{h}_{1}}\sqrt{k/r}$, $x={{h}_{2}}\sqrt{k/r}$ and $\alpha ={{r}_{0}}\theta /r$, in spite of a constant coefficient and an axial propagation phase, they are essentially the same thing.
When $\alpha =\pi /2$ and $-\pi /2$, Eq.(\ref{eq17}) regains the usual Fourier transforms, and it corresponds to a propagation distance $L=-r\pi /2$ and $r\pi /2$ on curved surface.
As a closely connection between the fractional Fourier transform and a parabolic index profile microlens\cite{RN47}, Wigner rotation\cite{RN32, RN30}, etc., it can now find a new correspondence as a curved surface.
As an example, the diffraction pattern of light field with a plane incidence is plotted in Fig.\ref{fig5}, the broken red line indicates an imaginary diagram. The revolution properties of transformation and diffraction pattern on curved surface are clearly revealed.

\section{conclusion}
We focused on the light beam propagation on two-dimensional curved surface. In this paper, through the formalization of eikonal and consequently the point spread function, a connection between the source contribution and the field distribution on curved surface was made. This connection is a generalized theory for an alternative ratio between the principle radius of surface and the wave length of light beam. Taking the Gaussian initial field into account, the evolution of not only the amplitude but also the phase is discussed. We find that the periodic boundary condition is contained naturally inside the geometric of surface of revolution. Another intriguing feature of fractional Fourier transformation is explored. This may provide more possibilities in laser optics and measurements in condensed matter and astronomical physics.

\section*{Funding}
This work was supported by the National Key Research and Development Program of China (Grant No. 2017YFC0601602).

\section*{Disclosures}
The authors declare no conflicts of interest.

\bibliography{curved_surface}

\begin{thebibliography}{31}%
\makeatletter
\providecommand \@ifxundefined [1]{%
 \@ifx{#1\undefined}
}%
\providecommand \@ifnum [1]{%
 \ifnum #1\expandafter \@firstoftwo
 \else \expandafter \@secondoftwo
 \fi
}%
\providecommand \@ifx [1]{%
 \ifx #1\expandafter \@firstoftwo
 \else \expandafter \@secondoftwo
 \fi
}%
\providecommand \natexlab [1]{#1}%
\providecommand \enquote  [1]{``#1''}%
\providecommand \bibnamefont  [1]{#1}%
\providecommand \bibfnamefont [1]{#1}%
\providecommand \citenamefont [1]{#1}%
\providecommand \href@noop [0]{\@secondoftwo}%
\providecommand \href [0]{\begingroup \@sanitize@url \@href}%
\providecommand \@href[1]{\@@startlink{#1}\@@href}%
\providecommand \@@href[1]{\endgroup#1\@@endlink}%
\providecommand \@sanitize@url [0]{\catcode `\\12\catcode `\$12\catcode
  `\&12\catcode `\#12\catcode `\^12\catcode `\_12\catcode `\%12\relax}%
\providecommand \@@startlink[1]{}%
\providecommand \@@endlink[0]{}%
\providecommand \url  [0]{\begingroup\@sanitize@url \@url }%
\providecommand \@url [1]{\endgroup\@href {#1}{\urlprefix }}%
\providecommand \urlprefix  [0]{URL }%
\providecommand \Eprint [0]{\href }%
\providecommand \doibase [0]{http://dx.doi.org/}%
\providecommand \selectlanguage [0]{\@gobble}%
\providecommand \bibinfo  [0]{\@secondoftwo}%
\providecommand \bibfield  [0]{\@secondoftwo}%
\providecommand \translation [1]{[#1]}%
\providecommand \BibitemOpen [0]{}%
\providecommand \bibitemStop [0]{}%
\providecommand \bibitemNoStop [0]{.\EOS\space}%
\providecommand \EOS [0]{\spacefactor3000\relax}%
\providecommand \BibitemShut  [1]{\csname bibitem#1\endcsname}%
\let\auto@bib@innerbib\@empty
\bibitem [{\citenamefont {Schultheiss}\ \emph {et~al.}(2015)\citenamefont
  {Schultheiss}, \citenamefont {Batz},\ and\ \citenamefont {Peschel}}]{RN15}%
  \BibitemOpen
  \bibfield  {author} {\bibinfo {author} {\bibfnamefont {V.~H.}\ \bibnamefont
  {Schultheiss}}, \bibinfo {author} {\bibfnamefont {S.}~\bibnamefont {Batz}}, \
  and\ \bibinfo {author} {\bibfnamefont {U.}~\bibnamefont {Peschel}},\ }\href
  {\doibase 10.1038/nphoton.2015.244} {\bibfield  {journal} {\bibinfo
  {journal} {Nature Photonics}\ }\textbf {\bibinfo {volume} {10}},\ \bibinfo
  {pages} {106} (\bibinfo {year} {2015})}\BibitemShut {NoStop}%
\bibitem [{\citenamefont {Brown}\ and\ \citenamefont {Twiss}(1956)}]{RN39}%
  \BibitemOpen
  \bibfield  {author} {\bibinfo {author} {\bibfnamefont {R.~H.}\ \bibnamefont
  {Brown}}\ and\ \bibinfo {author} {\bibfnamefont {R.~Q. J.~N.}\ \bibnamefont
  {Twiss}},\ }\href@noop {} {\bibfield  {journal} {\bibinfo  {journal}
  {Nature}\ }\textbf {\bibinfo {volume} {177}},\ \bibinfo {pages} {27}
  (\bibinfo {year} {1956})}\BibitemShut {NoStop}%
\bibitem [{\citenamefont {Xu}\ and\ \citenamefont {Wang}(2019)}]{RN24}%
  \BibitemOpen
  \bibfield  {author} {\bibinfo {author} {\bibfnamefont {C.}~\bibnamefont
  {Xu}}\ and\ \bibinfo {author} {\bibfnamefont {L.-G.}\ \bibnamefont {Wang}},\
  }\href {\doibase 10.1088/1367-2630/ab4f48} {\bibfield  {journal} {\bibinfo
  {journal} {New Journal of Physics}\ }\textbf {\bibinfo {volume} {21}}
  (\bibinfo {year} {2019}),\ 10.1088/1367-2630/ab4f48}\BibitemShut {NoStop}%
\bibitem [{\citenamefont {Libster-Hershko}\ \emph {et~al.}(2019)\citenamefont
  {Libster-Hershko}, \citenamefont {Shiloh},\ and\ \citenamefont
  {Arie}}]{RN25}%
  \BibitemOpen
  \bibfield  {author} {\bibinfo {author} {\bibfnamefont {A.}~\bibnamefont
  {Libster-Hershko}}, \bibinfo {author} {\bibfnamefont {R.}~\bibnamefont
  {Shiloh}}, \ and\ \bibinfo {author} {\bibfnamefont {A.}~\bibnamefont
  {Arie}},\ }\href {\doibase 10.1364/optica.6.000115} {\bibfield  {journal}
  {\bibinfo  {journal} {Optica}\ }\textbf {\bibinfo {volume} {6}} (\bibinfo
  {year} {2019}),\ 10.1364/optica.6.000115}\BibitemShut {NoStop}%
\bibitem [{\citenamefont {Tyc}\ \emph {et~al.}(2015)\citenamefont {Tyc},
  \citenamefont {Dao},\ and\ \citenamefont {Danner}}]{RN42}%
  \BibitemOpen
  \bibfield  {author} {\bibinfo {author} {\bibfnamefont {T.}~\bibnamefont
  {Tyc}}, \bibinfo {author} {\bibfnamefont {H.~L.}\ \bibnamefont {Dao}}, \ and\
  \bibinfo {author} {\bibfnamefont {A.~J.}\ \bibnamefont {Danner}},\ }\href
  {\doibase 10.1103/PhysRevA.92.053827} {\bibfield  {journal} {\bibinfo
  {journal} {Physical Review A}\ }\textbf {\bibinfo {volume} {92}} (\bibinfo
  {year} {2015}),\ 10.1103/PhysRevA.92.053827}\BibitemShut {NoStop}%
\bibitem [{\citenamefont {Xu}\ \emph {et~al.}(2019)\citenamefont {Xu},
  \citenamefont {Wang}, \citenamefont {Tyc}, \citenamefont {Sheng},
  \citenamefont {Zhu}, \citenamefont {Liu},\ and\ \citenamefont {Chen}}]{RN26}%
  \BibitemOpen
  \bibfield  {author} {\bibinfo {author} {\bibfnamefont {L.}~\bibnamefont
  {Xu}}, \bibinfo {author} {\bibfnamefont {X.}~\bibnamefont {Wang}}, \bibinfo
  {author} {\bibfnamefont {T.}~\bibnamefont {Tyc}}, \bibinfo {author}
  {\bibfnamefont {C.}~\bibnamefont {Sheng}}, \bibinfo {author} {\bibfnamefont
  {S.}~\bibnamefont {Zhu}}, \bibinfo {author} {\bibfnamefont {H.}~\bibnamefont
  {Liu}}, \ and\ \bibinfo {author} {\bibfnamefont {H.}~\bibnamefont {Chen}},\
  }\href {\doibase 10.1364/prj.7.001266} {\bibfield  {journal} {\bibinfo
  {journal} {Photonics Research}\ }\textbf {\bibinfo {volume} {7}} (\bibinfo
  {year} {2019}),\ 10.1364/prj.7.001266}\BibitemShut {NoStop}%
\bibitem [{\citenamefont {Tyc}\ \emph {et~al.}(2011)\citenamefont {Tyc},
  \citenamefont {Herzánová}, \citenamefont {Šarbort},\ and\ \citenamefont
  {Bering}}]{RN10}%
  \BibitemOpen
  \bibfield  {author} {\bibinfo {author} {\bibfnamefont {T.}~\bibnamefont
  {Tyc}}, \bibinfo {author} {\bibfnamefont {L.}~\bibnamefont {Herzánová}},
  \bibinfo {author} {\bibfnamefont {M.}~\bibnamefont {Šarbort}}, \ and\
  \bibinfo {author} {\bibfnamefont {K.}~\bibnamefont {Bering}},\ }\href
  {\doibase 10.1088/1367-2630/13/11/115004} {\bibfield  {journal} {\bibinfo
  {journal} {New Journal of Physics}\ }\textbf {\bibinfo {volume} {13}}
  (\bibinfo {year} {2011}),\ 10.1088/1367-2630/13/11/115004}\BibitemShut
  {NoStop}%
\bibitem [{\citenamefont {Xu}\ and\ \citenamefont {Chen}(2014)}]{RN14}%
  \BibitemOpen
  \bibfield  {author} {\bibinfo {author} {\bibfnamefont {L.}~\bibnamefont
  {Xu}}\ and\ \bibinfo {author} {\bibfnamefont {H.}~\bibnamefont {Chen}},\
  }\href {\doibase 10.1038/nphoton.2014.307} {\bibfield  {journal} {\bibinfo
  {journal} {Nature Photonics}\ }\textbf {\bibinfo {volume} {9}},\ \bibinfo
  {pages} {15} (\bibinfo {year} {2014})}\BibitemShut {NoStop}%
\bibitem [{\citenamefont {Šarbort}\ and\ \citenamefont {Tyc}(2012)}]{RN11}%
  \BibitemOpen
  \bibfield  {author} {\bibinfo {author} {\bibfnamefont {M.}~\bibnamefont
  {Šarbort}}\ and\ \bibinfo {author} {\bibfnamefont {T.}~\bibnamefont {Tyc}},\
  }\href {\doibase 10.1088/2040-8978/14/7/075705} {\bibfield  {journal}
  {\bibinfo  {journal} {Journal of Optics}\ }\textbf {\bibinfo {volume} {14}}
  (\bibinfo {year} {2012}),\ 10.1088/2040-8978/14/7/075705}\BibitemShut
  {NoStop}%
\bibitem [{\citenamefont {Leonhardt}(2006)}]{RN2}%
  \BibitemOpen
  \bibfield  {author} {\bibinfo {author} {\bibfnamefont {U.}~\bibnamefont
  {Leonhardt}},\ }\href {\doibase 10.1126/science.1126493} {\bibfield
  {journal} {\bibinfo  {journal} {Science}\ }\textbf {\bibinfo {volume}
  {312}},\ \bibinfo {pages} {1777} (\bibinfo {year} {2006})}\BibitemShut
  {NoStop}%
\bibitem [{\citenamefont {Xu}\ \emph {et~al.}(2020)\citenamefont {Xu},
  \citenamefont {Xiao}, \citenamefont {Zhang}, \citenamefont {Li},
  \citenamefont {Zhou},\ and\ \citenamefont {Chen}}]{RN28}%
  \BibitemOpen
  \bibfield  {author} {\bibinfo {author} {\bibfnamefont {L.}~\bibnamefont
  {Xu}}, \bibinfo {author} {\bibfnamefont {W.}~\bibnamefont {Xiao}}, \bibinfo
  {author} {\bibfnamefont {L.}~\bibnamefont {Zhang}}, \bibinfo {author}
  {\bibfnamefont {J.}~\bibnamefont {Li}}, \bibinfo {author} {\bibfnamefont
  {J.}~\bibnamefont {Zhou}}, \ and\ \bibinfo {author} {\bibfnamefont
  {H.}~\bibnamefont {Chen}},\ }\href {\doibase 10.1364/OE.395351} {\bibfield
  {journal} {\bibinfo  {journal} {Opt Express}\ }\textbf {\bibinfo {volume}
  {28}},\ \bibinfo {pages} {20215} (\bibinfo {year} {2020})}\BibitemShut
  {NoStop}%
\bibitem [{\citenamefont {Batz}\ and\ \citenamefont {Peschel}(2010)}]{RN7}%
  \BibitemOpen
  \bibfield  {author} {\bibinfo {author} {\bibfnamefont {S.}~\bibnamefont
  {Batz}}\ and\ \bibinfo {author} {\bibfnamefont {U.}~\bibnamefont {Peschel}},\
  }\href {\doibase 10.1103/PhysRevA.81.053806} {\bibfield  {journal} {\bibinfo
  {journal} {Physical Review A}\ }\textbf {\bibinfo {volume} {81}} (\bibinfo
  {year} {2010}),\ 10.1103/PhysRevA.81.053806}\BibitemShut {NoStop}%
\bibitem [{\citenamefont {Conti}(2016)}]{RN16}%
  \BibitemOpen
  \bibfield  {author} {\bibinfo {author} {\bibfnamefont {C.}~\bibnamefont
  {Conti}},\ }\href {\doibase 10.1007/s11434-016-1040-z} {\bibfield  {journal}
  {\bibinfo  {journal} {Science Bulletin}\ }\textbf {\bibinfo {volume} {61}},\
  \bibinfo {pages} {570} (\bibinfo {year} {2016})}\BibitemShut {NoStop}%
\bibitem [{\citenamefont {Perczel}\ \emph {et~al.}(2018)\citenamefont
  {Perczel}, \citenamefont {Kómár},\ and\ \citenamefont {Lukin}}]{RN20}%
  \BibitemOpen
  \bibfield  {author} {\bibinfo {author} {\bibfnamefont {J.}~\bibnamefont
  {Perczel}}, \bibinfo {author} {\bibfnamefont {P.}~\bibnamefont {Kómár}}, \
  and\ \bibinfo {author} {\bibfnamefont {M.~D.}\ \bibnamefont {Lukin}},\ }\href
  {\doibase 10.1103/PhysRevA.98.033803} {\bibfield  {journal} {\bibinfo
  {journal} {Physical Review A}\ }\textbf {\bibinfo {volume} {98}} (\bibinfo
  {year} {2018}),\ 10.1103/PhysRevA.98.033803}\BibitemShut {NoStop}%
\bibitem [{\citenamefont {Schultheiss}\ \emph {et~al.}(2020)\citenamefont
  {Schultheiss}, \citenamefont {Batz},\ and\ \citenamefont {Peschel}}]{RN27}%
  \BibitemOpen
  \bibfield  {author} {\bibinfo {author} {\bibfnamefont {V.~H.}\ \bibnamefont
  {Schultheiss}}, \bibinfo {author} {\bibfnamefont {S.}~\bibnamefont {Batz}}, \
  and\ \bibinfo {author} {\bibfnamefont {U.}~\bibnamefont {Peschel}},\ }\href
  {\doibase 10.1080/23746149.2020.1759451} {\bibfield  {journal} {\bibinfo
  {journal} {Advances in Physics: X}\ }\textbf {\bibinfo {volume} {5}}
  (\bibinfo {year} {2020}),\ 10.1080/23746149.2020.1759451}\BibitemShut
  {NoStop}%
\bibitem [{\citenamefont {Batz}\ and\ \citenamefont {Peschel}(2008)}]{RN4}%
  \BibitemOpen
  \bibfield  {author} {\bibinfo {author} {\bibfnamefont {S.}~\bibnamefont
  {Batz}}\ and\ \bibinfo {author} {\bibfnamefont {U.}~\bibnamefont {Peschel}},\
  }\href {\doibase 10.1103/PhysRevA.78.043821} {\bibfield  {journal} {\bibinfo
  {journal} {Physical Review A}\ }\textbf {\bibinfo {volume} {78}} (\bibinfo
  {year} {2008}),\ 10.1103/PhysRevA.78.043821}\BibitemShut {NoStop}%
\bibitem [{\citenamefont {Tyc}(2013)}]{RN13}%
  \BibitemOpen
  \bibfield  {author} {\bibinfo {author} {\bibfnamefont {T.}~\bibnamefont
  {Tyc}},\ }\href {\doibase 10.1088/1367-2630/15/6/065005} {\bibfield
  {journal} {\bibinfo  {journal} {New Journal of Physics}\ }\textbf {\bibinfo
  {volume} {15}} (\bibinfo {year} {2013}),\
  10.1088/1367-2630/15/6/065005}\BibitemShut {NoStop}%
\bibitem [{\citenamefont {Lai}\ \emph {et~al.}(2018)\citenamefont {Lai},
  \citenamefont {Wang}, \citenamefont {Liang}, \citenamefont {Wang},\ and\
  \citenamefont {Zong}}]{RN19}%
  \BibitemOpen
  \bibfield  {author} {\bibinfo {author} {\bibfnamefont {M.-Y.}\ \bibnamefont
  {Lai}}, \bibinfo {author} {\bibfnamefont {Y.-L.}\ \bibnamefont {Wang}},
  \bibinfo {author} {\bibfnamefont {G.-H.}\ \bibnamefont {Liang}}, \bibinfo
  {author} {\bibfnamefont {F.}~\bibnamefont {Wang}}, \ and\ \bibinfo {author}
  {\bibfnamefont {H.-S.}\ \bibnamefont {Zong}},\ }\href {\doibase
  10.1103/PhysRevA.97.033843} {\bibfield  {journal} {\bibinfo  {journal}
  {Physical Review A}\ }\textbf {\bibinfo {volume} {97}} (\bibinfo {year}
  {2018}),\ 10.1103/PhysRevA.97.033843}\BibitemShut {NoStop}%
\bibitem [{\citenamefont {Wang}\ \emph {et~al.}(2018)\citenamefont {Wang},
  \citenamefont {Lai}, \citenamefont {Wang}, \citenamefont {Zong},\ and\
  \citenamefont {Chen}}]{RN21}%
  \BibitemOpen
  \bibfield  {author} {\bibinfo {author} {\bibfnamefont {Y.-L.}\ \bibnamefont
  {Wang}}, \bibinfo {author} {\bibfnamefont {M.-Y.}\ \bibnamefont {Lai}},
  \bibinfo {author} {\bibfnamefont {F.}~\bibnamefont {Wang}}, \bibinfo {author}
  {\bibfnamefont {H.-S.}\ \bibnamefont {Zong}}, \ and\ \bibinfo {author}
  {\bibfnamefont {Y.-F.}\ \bibnamefont {Chen}},\ }\href {\doibase
  10.1103/PhysRevA.97.042108} {\bibfield  {journal} {\bibinfo  {journal}
  {Physical Review A}\ }\textbf {\bibinfo {volume} {97}} (\bibinfo {year}
  {2018}),\ 10.1103/PhysRevA.97.042108}\BibitemShut {NoStop}%
\bibitem [{\citenamefont {Ferrari}\ and\ \citenamefont {Cuoghi}(2008)}]{RN5}%
  \BibitemOpen
  \bibfield  {author} {\bibinfo {author} {\bibfnamefont {G.}~\bibnamefont
  {Ferrari}}\ and\ \bibinfo {author} {\bibfnamefont {G.}~\bibnamefont
  {Cuoghi}},\ }\href {\doibase 10.1103/PhysRevLett.100.230403} {\bibfield
  {journal} {\bibinfo  {journal} {Phys Rev Lett}\ }\textbf {\bibinfo {volume}
  {100}},\ \bibinfo {pages} {230403} (\bibinfo {year} {2008})}\BibitemShut
  {NoStop}%
\bibitem [{\citenamefont {Schultheiss}\ \emph {et~al.}(2010)\citenamefont
  {Schultheiss}, \citenamefont {Batz}, \citenamefont {Szameit}, \citenamefont
  {Dreisow}, \citenamefont {Nolte}, \citenamefont {Tunnermann}, \citenamefont
  {Longhi},\ and\ \citenamefont {Peschel}}]{RN8}%
  \BibitemOpen
  \bibfield  {author} {\bibinfo {author} {\bibfnamefont {V.~H.}\ \bibnamefont
  {Schultheiss}}, \bibinfo {author} {\bibfnamefont {S.}~\bibnamefont {Batz}},
  \bibinfo {author} {\bibfnamefont {A.}~\bibnamefont {Szameit}}, \bibinfo
  {author} {\bibfnamefont {F.}~\bibnamefont {Dreisow}}, \bibinfo {author}
  {\bibfnamefont {S.}~\bibnamefont {Nolte}}, \bibinfo {author} {\bibfnamefont
  {A.}~\bibnamefont {Tunnermann}}, \bibinfo {author} {\bibfnamefont
  {S.}~\bibnamefont {Longhi}}, \ and\ \bibinfo {author} {\bibfnamefont
  {U.}~\bibnamefont {Peschel}},\ }\href {\doibase
  10.1103/PhysRevLett.105.143901} {\bibfield  {journal} {\bibinfo  {journal}
  {Phys Rev Lett}\ }\textbf {\bibinfo {volume} {105}},\ \bibinfo {pages}
  {143901} (\bibinfo {year} {2010})}\BibitemShut {NoStop}%
\bibitem [{\citenamefont {Patsyk}\ \emph {et~al.}(2018)\citenamefont {Patsyk},
  \citenamefont {Bandres}, \citenamefont {Bekenstein},\ and\ \citenamefont
  {Segev}}]{RN23}%
  \BibitemOpen
  \bibfield  {author} {\bibinfo {author} {\bibfnamefont {A.}~\bibnamefont
  {Patsyk}}, \bibinfo {author} {\bibfnamefont {M.~A.}\ \bibnamefont {Bandres}},
  \bibinfo {author} {\bibfnamefont {R.}~\bibnamefont {Bekenstein}}, \ and\
  \bibinfo {author} {\bibfnamefont {M.}~\bibnamefont {Segev}},\ }\href
  {\doibase 10.1103/PhysRevX.8.011001} {\bibfield  {journal} {\bibinfo
  {journal} {Physical Review X}\ }\textbf {\bibinfo {volume} {8}} (\bibinfo
  {year} {2018}),\ 10.1103/PhysRevX.8.011001}\BibitemShut {NoStop}%
\bibitem [{\citenamefont {Wang}\ \emph {et~al.}(2017)\citenamefont {Wang},
  \citenamefont {Chen}, \citenamefont {Liu}, \citenamefont {Xu}, \citenamefont
  {Sheng},\ and\ \citenamefont {Zhu}}]{RN43}%
  \BibitemOpen
  \bibfield  {author} {\bibinfo {author} {\bibfnamefont {X.}~\bibnamefont
  {Wang}}, \bibinfo {author} {\bibfnamefont {H.}~\bibnamefont {Chen}}, \bibinfo
  {author} {\bibfnamefont {H.}~\bibnamefont {Liu}}, \bibinfo {author}
  {\bibfnamefont {L.}~\bibnamefont {Xu}}, \bibinfo {author} {\bibfnamefont
  {C.}~\bibnamefont {Sheng}}, \ and\ \bibinfo {author} {\bibfnamefont
  {S.}~\bibnamefont {Zhu}},\ }\href {\doibase 10.1103/PhysRevLett.119.033902}
  {\bibfield  {journal} {\bibinfo  {journal} {Phys Rev Lett}\ }\textbf
  {\bibinfo {volume} {119}},\ \bibinfo {pages} {033902} (\bibinfo {year}
  {2017})}\BibitemShut {NoStop}%
\bibitem [{\citenamefont {Jeffreys}(1925)}]{RN44}%
  \BibitemOpen
  \bibfield  {author} {\bibinfo {author} {\bibfnamefont {H.}~\bibnamefont
  {Jeffreys}},\ }\href {\doibase 10.1112/plms/s2-23.1.428} {\bibfield
  {journal} {\bibinfo  {journal} {Proceedings of the London Mathematical
  Society}\ }\textbf {\bibinfo {volume} {s2-23}},\ \bibinfo {pages} {428}
  (\bibinfo {year} {1925})}\BibitemShut {NoStop}%
\bibitem [{\citenamefont {Wald}(2010)}]{RN45}%
  \BibitemOpen
  \bibfield  {author} {\bibinfo {author} {\bibfnamefont {R.~M.}\ \bibnamefont
  {Wald}},\ }\href@noop {} {\emph {\bibinfo {title} {GENERAL RELATIVITY}}}\
  (\bibinfo  {publisher} {University of Chicago press},\ \bibinfo {year}
  {2010})\BibitemShut {NoStop}%
\bibitem [{\citenamefont {da~Costa}(1981)}]{RN1}%
  \BibitemOpen
  \bibfield  {author} {\bibinfo {author} {\bibfnamefont {R.~C.~T.}\
  \bibnamefont {da~Costa}},\ }\href {\doibase 10.1103/PhysRevA.23.1982}
  {\bibfield  {journal} {\bibinfo  {journal} {Physical Review A}\ }\textbf
  {\bibinfo {volume} {23}},\ \bibinfo {pages} {1982} (\bibinfo {year}
  {1981})}\BibitemShut {NoStop}%
\bibitem [{\citenamefont {Andresen}\ \emph {et~al.}(2013)\citenamefont
  {Andresen}, \citenamefont {Finot}, \citenamefont {Oron},\ and\ \citenamefont
  {Rigneault}}]{RN46}%
  \BibitemOpen
  \bibfield  {author} {\bibinfo {author} {\bibfnamefont {E.~R.}\ \bibnamefont
  {Andresen}}, \bibinfo {author} {\bibfnamefont {C.}~\bibnamefont {Finot}},
  \bibinfo {author} {\bibfnamefont {D.}~\bibnamefont {Oron}}, \ and\ \bibinfo
  {author} {\bibfnamefont {H.}~\bibnamefont {Rigneault}},\ }\href {\doibase
  10.1103/PhysRevLett.110.143902} {\bibfield  {journal} {\bibinfo  {journal}
  {Phys Rev Lett}\ }\textbf {\bibinfo {volume} {110}},\ \bibinfo {pages}
  {143902} (\bibinfo {year} {2013})}\BibitemShut {NoStop}%
\bibitem [{\citenamefont {Namias}(1980)}]{RN29}%
  \BibitemOpen
  \bibfield  {author} {\bibinfo {author} {\bibfnamefont {V.}~\bibnamefont
  {Namias}},\ }\href {\doibase 10.1093/imamat/25.3.241} {\bibfield  {journal}
  {\bibinfo  {journal} {IMA Journal of Applied Mathematics}\ }\textbf {\bibinfo
  {volume} {25}},\ \bibinfo {pages} {241} (\bibinfo {year} {1980})}\BibitemShut
  {NoStop}%
\bibitem [{\citenamefont {Dragoman}\ \emph {et~al.}(1999)\citenamefont
  {Dragoman}, \citenamefont {Dragoman},\ and\ \citenamefont {Brenner}}]{RN47}%
  \BibitemOpen
  \bibfield  {author} {\bibinfo {author} {\bibfnamefont {D.}~\bibnamefont
  {Dragoman}}, \bibinfo {author} {\bibfnamefont {M.}~\bibnamefont {Dragoman}},
  \ and\ \bibinfo {author} {\bibfnamefont {K.~H.}\ \bibnamefont {Brenner}},\
  }\href {\doibase 10.1364/ol.24.000933} {\bibfield  {journal} {\bibinfo
  {journal} {Opt Lett}\ }\textbf {\bibinfo {volume} {24}},\ \bibinfo {pages}
  {933} (\bibinfo {year} {1999})}\BibitemShut {NoStop}%
\bibitem [{\citenamefont {Lohmann}(1993)}]{RN32}%
  \BibitemOpen
  \bibfield  {author} {\bibinfo {author} {\bibfnamefont {A.~W.}\ \bibnamefont
  {Lohmann}},\ }\href {\doibase 10.1364/josaa.10.002181} {\bibfield  {journal}
  {\bibinfo  {journal} {Journal of the Optical Society of America A}\ }\textbf
  {\bibinfo {volume} {10}} (\bibinfo {year} {1993}),\
  10.1364/josaa.10.002181}\BibitemShut {NoStop}%
\bibitem [{\citenamefont {Mendlovic}\ and\ \citenamefont
  {Ozaktas}(1993)}]{RN30}%
  \BibitemOpen
  \bibfield  {author} {\bibinfo {author} {\bibfnamefont {D.}~\bibnamefont
  {Mendlovic}}\ and\ \bibinfo {author} {\bibfnamefont {H.~M.}\ \bibnamefont
  {Ozaktas}},\ }\href {\doibase 10.1364/josaa.10.001875} {\bibfield  {journal}
  {\bibinfo  {journal} {Journal of the Optical Society of America A}\ }\textbf
  {\bibinfo {volume} {10}} (\bibinfo {year} {1993}),\
  10.1364/josaa.10.001875}\BibitemShut {NoStop}%
\end{thebibliography}%

\end{document}